\documentclass[aps,prc,reprint,nofootinbib]{revtex4-1}
\usepackage{graphicx}   
\usepackage{latexsym}   
\usepackage{enumerate}
\usepackage{rotating,booktabs,multirow}

\begin{document}
\title{Deuteron production from phase-space coalescence in the UrQMD approach}
\author{Sukanya Sombun$^{1}$,  Kristiya Tomuang$^{2}$, Ayut Limphirat$^{1}$, Paula~Hillmann$^{3,4}$, Christoph Herold$^{1}$, Jan~Steinheimer$^3$, Yupeng Yan$^{1}$,  Marcus~Bleicher$^{3,4,5,6}$}

\affiliation{$^1$ School of Physics and Center of Excellence in High Energy Physics $\&$ Astrophysics, Suranaree University of Technology, Nakhon Ratchasima 30000, Thailand}
\affiliation{$^2$ Department of Physics, Naresuan University, Phitsanulok 65000, Thailand}
\affiliation{$^3$ Frankfurt Institute for Advanced Studies, Ruth-Moufang-Str. 1, 60438 Frankfurt am Main, Germany}
\affiliation{$^4$ Institut f\"ur Theoretische Physik, Goethe Universit\"at Frankfurt, Max-von-Laue-Strasse 1, D-60438 Frankfurt am Main, Germany}
\affiliation{$^5$ GSI Helmholtzzentrum f\"ur Schwerionenforschung GmbH, Planckstr. 1, 64291 Darmstadt , Germany}
\affiliation{$^6$ John von Neumann-Institut f\"ur Computing, Forschungzentrum J\"ulich,
52425 J\"ulich, Germany}

\begin{abstract}
UrQMD phase-space coalescence calculations for the production of deuterons are compared with available data for various reactions from the GSI/FAIR energy regime over the CERN-SPS/RHIC-BES region up to LHC energies. It is found that the production process of deuterons, as reflected in their rapidity and transverse momentum distributions in p+p, p+A and A+A collisions in the mentioned energy regime are in good agreement with experimental data. We further explore the energy and centrality dependence of the d/p and $\bar{\mathrm{d}}/\bar{\mathrm{p}}$ ratios and compare with thermal model results. Finally, we discuss anti-deuteron production for selected systems. Overall, a good description of the experimental data is observed. Most importantly this good description is based only on a single set of coalescence parameters that is independent of energy and system size and can also be applied for anti-deuterons.
\end{abstract}

\maketitle
\section{Introduction}

The collision of heavy and light ions at various beam energies allows to explore the properties and dynamics of strongly interacting matter, i.e. matter governed by the laws of Quantum Chromodynamics, QCD. QCD matter under extreme conditions has been present during the first microseconds after the Big Bang and is nowadays present in neutron star mergers and other compact stellar objects. An ideal environment to probe dense QCD matter in the laboratory is given by the collision of light and heavy ions in accelerators like the SIS18 at GSI or at the future FAIR facility. Here similar energies, as previously in collisions at the BNL-AGS and the CERN-SPS (especially in the beam energy scan run by experiment NA49) are probed. Nowadays systematic studies continue with the running programs at the CERN-SPS (NA61 experiment) and at RHIC-BES (STAR experiment). The production of nuclear clusters, e.g. deuterons, is especially interesting because it can shed light on the formation process, e.g. direct thermal production or coalescence. This is especially interesting in the light of the recent ALICE data (on nuclear clusters) at LHC and their interpretation in terms of the thermal model \cite{Andronic:2012dm}. At high beam energies it may be difficult to distinguish coalescence from direct thermal production from the yields alone, as the yields of both processes seem to be similar \cite{Mrowczynski:2016xqm}. Here different clusters, e.g. the anti-hyper-triton may provide additional insights. On the other hand, differential observables like $p_t$ distributions and collective flow may turn out to be more sensitive to the formation process and time. In addition, deuteron formation by coalescence provides information on the 2-particle phase space distribution of nucleons complementary to HBT studies~\cite{Mrowczynski:1992gc}. To study deuteron formation in a wide energy range, we employ the UrQMD model \cite{Bass:1998ca,Bleicher:1999xi} to investigate deuteron production from phase-space coalescence. This will allow us to probe the phase space density of nucleons at kinetic freeze-out and therefore constraints the evolution of the system. In the energy regime from $E_{lab}=1$A GeV - 160A GeV a multitude of experimental data on cluster production is available. The measured yields and spectra ($dN/dy$ and $dN/dm_t$) of deuterons and protons are compared to experimental data. Moreover, the energy dependence of the ratio of (anti-)deuterons to (anti-)protons in p+p and nucleus-nucleus collisions are calculated and compared with experimental data and in the case of A+A collisions with thermal model fits.

\subsection{The UrQMD model with coalescence}
The Ultra relativistic Quantum Molecular Dynamics (UrQMD) transport model is based on  the binary elastic and inelastic scattering of hadrons, including resonance excitations and decays, as well as string dynamics and strangeness exchange reactions \cite{Bass:1998ca,Bleicher:1999xi,Graef:2014mra}. The model employs a geometrical interpretation of the scattering cross sections which are taken, when available, from experimental data \cite{Olive:2016xmw} or model calculations.
	
The default version of UrQMD does not include the formation of  deuterons or other nuclear clusters. To calculate the abundances and spectra of nuclear clusters different approaches are possible. The different approaches discussed below are related to the original quantum-mechanical momentum-space coalescence model as formulated by Sato and Yazaki \cite{Sato:1981ez}. Sato and Yazaki calculate the deuteron formation probability from the spatially integrated transition amplitude of the n-p density matrix (assumed to factorize in independent single particle proton and neutron densities, space-momentum correlations are neglected) with the deuteron wave function. Protons and neutrons with momenta $k\pm \Delta p$ ($k$ being the deuteron momentum) do then coalesce into the deuteron state with the quantity $\Delta p$ being related to the deuteron wave function, given a certain spatial distribution of protons and neutrons. Folding the deuteron wave function with the spatial distribution of the n-p source allows then to introduce a single momentum space parameter $p_0$. If the wave function is small compared to the source size, $p_0$ is inversely proportional to the source volume. Therefore it is clear that $p_0$ encodes also information on the emission source in this approach and is to first order system size ($1/{\rm volume}$) dependent.
\begin{figure}[t]	
\includegraphics[width=0.5\textwidth]{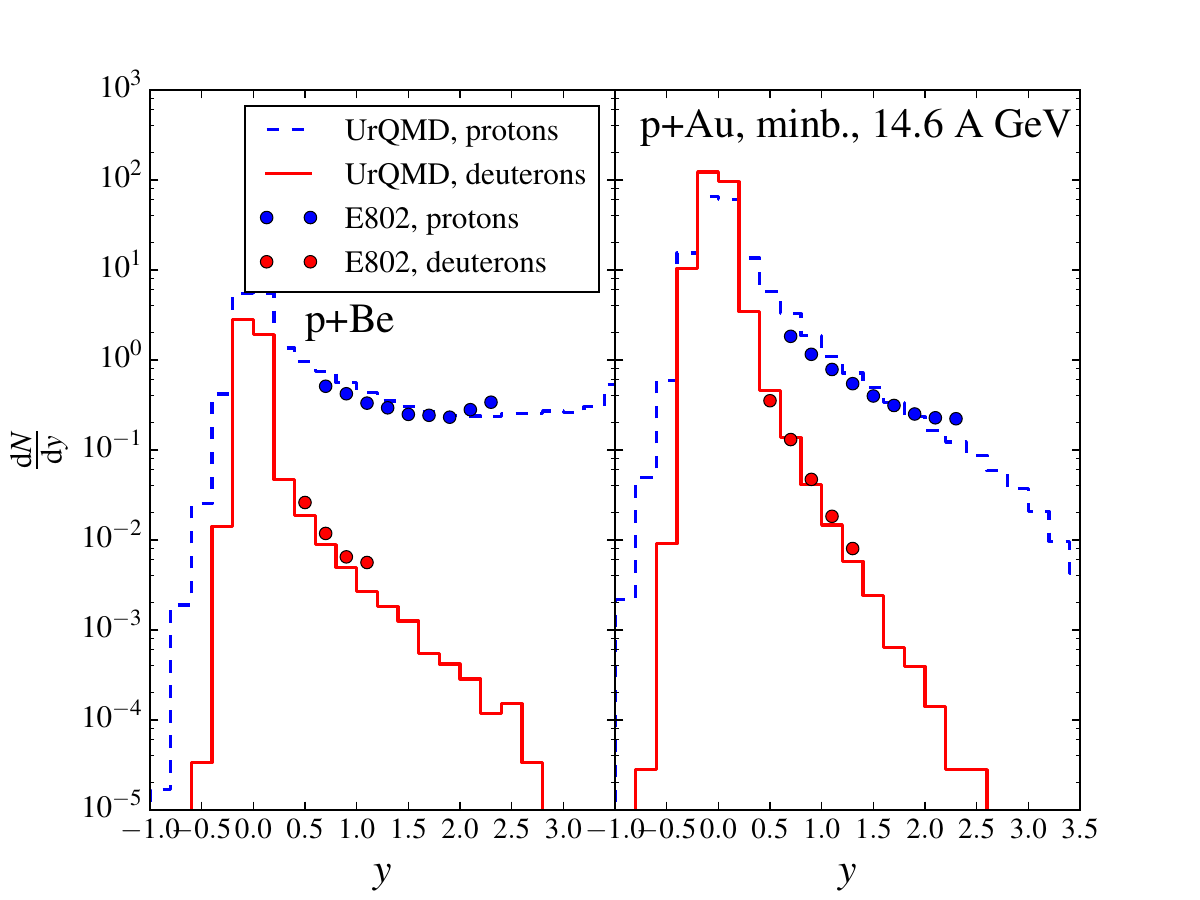}
\caption{[Color online] Rapidity distributions of protons and deuterons in minimum bias p+Be (left) and p+Au (right) collisions at a beam energy of $14.6$A GeV, from the UrQMD model (lines) compared to experimental E802 data (symbols) \cite{Abbott:1991en}.
}\label{f1}
\end{figure}		

In previous calculations using the UrQMD hybrid approach \cite{Steinheimer:2012tb} the production of clusters was calculated via the Cooper-Frye equation on a hyper-surface of constant energy density. This approach assumes that the deuterons are not formed by coalescence, but are emitted as a single entity from the fireball as suggested in statistical hadronization models. An alternative way is the coalescence approach introduced by Gyulassy, Frankel, and Remler\cite{Gyulassy:1982pe} based on the von Neumann equation for the n-body density. This "Wigner function" approach follows in spirit the original idea by Sato and Yazaki, but suggests to project the Wigner-transformed wave function on the classical phase space distribution generated from simulations, under the assumption that the classical phase space density provides a good approximation of the (factorized) n-p density matrix. The main advantage in this approach is that one does not need to integrate the spatial volume of the source into the coalescence parameter, but uses the relative space-momentum dependent Wigner representation of the deuteron state directly. Here one can also easily include the space-momentum correlations of the protons/neutrons emerging during the reaction. The Wigner function approach has been applied very successfully in the description of deuteron production, see e.g. \cite{Aichelin:1987rh,Nagle:1994wj,Nagle:1996vp,Ko:2010zza,Zhu:2015voa}.
\begin{figure}[t]	
	\includegraphics[width=0.5\textwidth]{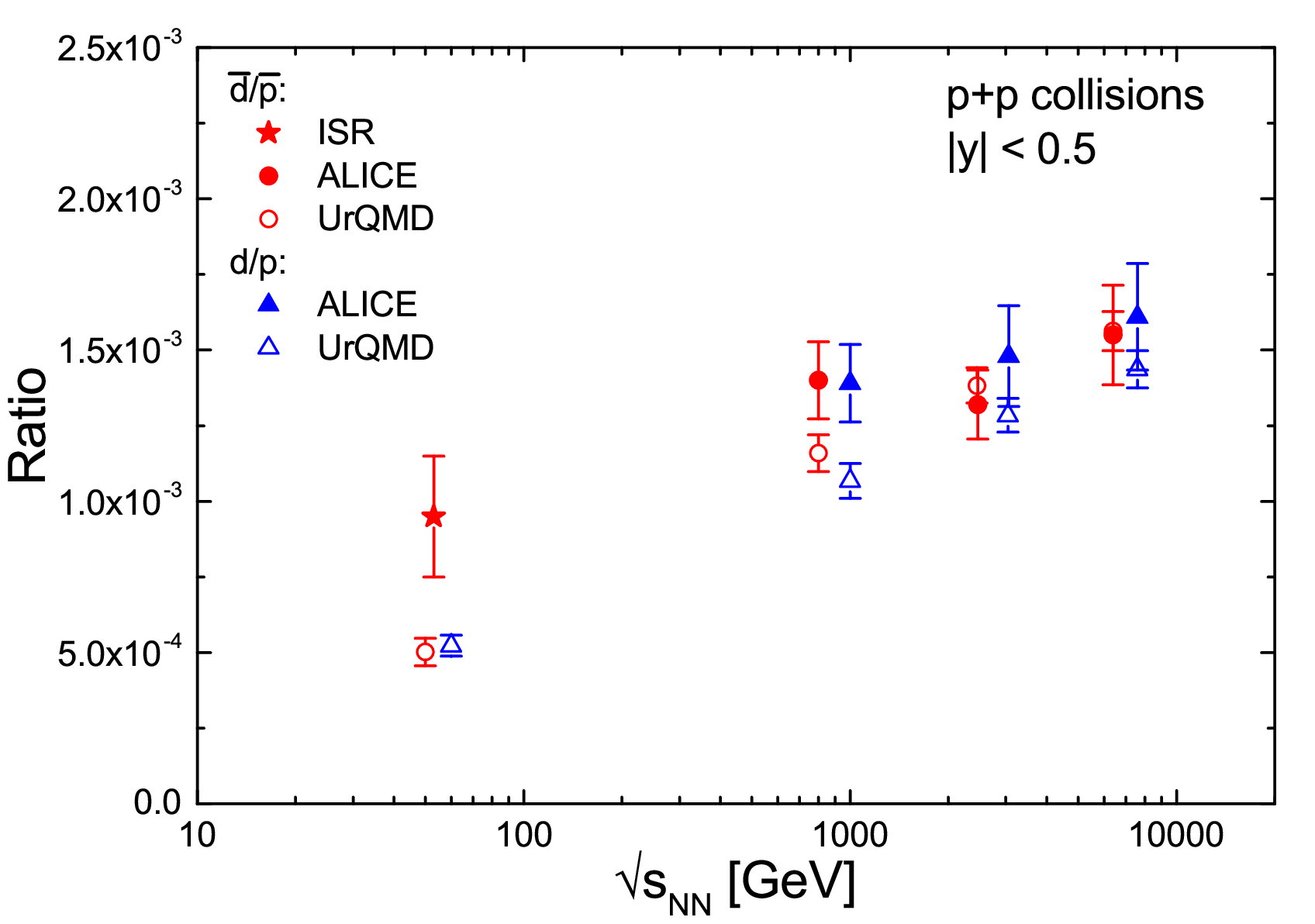}
	\caption{[Color online] Energy dependence of d/p and $ \overline{d} $/ $ \overline{p} $ ratios in pp collisions with $|y|<0.5$ at $\sqrt{s_{NN}}= 53, 900, 2760$ and $7000$~GeV. The open symbols represent UrQMD model results. The solid symbols denote the result from ISR (star) \cite{Alper:1973my,Henning:1977mt,Alper:1975jm} and ALICE (circle and triangle) \cite{Acharya:2017fvb}}.
    \label{f2} 
\end{figure}		

Another well tested possibility is to use a cut-off coalescence approach \cite{Li:2015pta}, either in momentum space or coordinate space or in full phase space. This approach is similar to the Wigner function approach, but essentially assumes a flat probability in coordinate space and momentum space for the coalescence probability (instead of the deuteron wave function). One defines a maximum relative momentum $ \Delta p$ and/or a maximum distance $\Delta r$ between the proton and the neutron to form a deuteron. If one restricts oneself to the relative momentum cut only, one observes a similar volume dependence of the momentum space coalescence parameter as in the Sato/Yazaki approach. As in the Wigner function approach, the inclusion of a space and momentum space parameter allows to use a volume independent set of parameters. Phase space coalescence has been shown to work successfully and to yield results similar to the Wigner function approach, see e.g. \cite{Nagle:1996vp}.

For the purpose of this work, we model deuteron formation in UrQMD via phase space coalescence at the point of last interaction of the respective proton and neutron in space and time. 
The method we use comprises the following steps:
\begin{enumerate}
\item During the evolution of the system, we follow the protons and neutrons until their individual space-time points of last interaction. 
\item For each p-n pair, the momentum and position of proton and neutron is boosted to the 2-particle  rest-frame of this p-n pair.
\item The particle that has decoupled earlier is then propagated to the later time of the other particle. 
\item We calculate the relative momenta  $ \Delta p=|\overrightarrow{p}_{1}-\overrightarrow{p}_{2}|$ and the relative distances $ \Delta r =|\overrightarrow{x}_{1}-\overrightarrow{x}_{2}|$ of the p-n pair in the 2-particle rest-frame at equal times. The yield of deuteron candidates is then given by the condition of $ \Delta p< \Delta p_{max}$ and $ \Delta r< \Delta r_{max}$. Here we use the parameter set of $ \Delta p_{max}=0.285 $ GeV/c and $ \Delta r_{max} =3.575$ fm. 
\item For each deuteron candidate we perform the statistical spin and isospin projection\footnote{%
The statistical spin and isospin factors emerge from the summation and averaging over spin states and from the condition of anti-symmetry of the deuteron wave function. The deuteron state itself has the quantum numbers $S=1$, and $I=0$.
As shown in detail in \cite{Xia:2014rua}, the direct product of the spin- and isospin wave-functions of a proton and a neutron generates eight combinations of n-p wave functions. However, only three states belong to the deuteron quantum numbers ($I=0$, $S=1$, $S_z$=-1, 0, 1). Thus, one obtains a statistical spin and isospin factor of $3/8$. 
}%
to the deuteron state (probability $1/2\cdot3/4 = 3/8$) \cite{Nagle:1996vp,Xia:2014rua}. Then, the  chosen p-n pair is marked as a deuteron and its constituent nucleons are removed from the phase space distribution. 
\end{enumerate}

It is important to note that the parameters for deuteron formation are kept independent of energy, collision system and centrality, because they are related to the deuteron wave function. As we will see, the chosen parameter values provide a good description of the available data in a wide range of systems and beam energies. 

\section{Results} 
In the following we will present extensive comparisons of UrQMD model results with experimental measurements of deuteron production at various beam energies and system sizes. We will mainly distinguish between proton induced reactions, p+p and p+A and nuclear reactions A+A. 
The calculated yields, ratios, rapidity and transverse momentum distributions will give us good insights into the validity of the coalescence approach and possible shortcomings. For Pb+Pb collisions of 2.76 TeV, UrQMD is used in hybrid mode.

All simulations are performed using UrQMD with deuteron production via the coalescence approach as described above. 

\subsection{Proton induced reactions}
Proton-proton and proton-nucleus reactions provide the simplest test cases for our model studies. In these systems the rescattering stage is rather short and the freeze-out volumes are smaller than in nucleus-nucleus reactions. In comparison to the following nucleus-nucleus studies, it provides a handle to explore the independence of the coalescence parameters on the system size. 

Figure \ref{f1} shows the rapidity distributions of protons and deuterons in minimum bias p+Au (left) and p+Be (right) collisions at a beam energy of $14.6$A GeV. The symbols denote the experimental data, the lines indicate the UrQMD calculations. The deuteron and proton yields are consistent with the experimental E802 data \cite{Abbott:1991en}, and the rapidity distributions are well reproduced.

A similarly good description of the deuteron and anti-deuteron production in proton-proton reactions can also be obtained for the highest beam energies achievable at the LHC.
The integrated, midrapidity $|y|<0.5$, yields ($dN/dy$) of deuterons and anti-deuterons in p+p collisions are calculated by the UrQMD model for different center-of-mass energies $\sqrt{s_{NN}}= 0.9, 2.76$ and $7$ TeV and compared to recent ALICE data, as shown in Table I. We can see that our results are in agreement with the ALICE experimental data.

\begin{table}[t]
		\centering		
		\label{}
       \begin{tabular}{p{0.5cm}p{1cm}p{3.5cm}p{3cm}}
			
			\hline
			&  \multirow{1}[3]{*}{$ \sqrt{s_{NN}} $(TeV)}  & \multicolumn{2}{c}{$ dN/dy $}   \\ \cmidrule(rl){3-4} 
			&  & \multicolumn{1}{c} {ALICE} & \multicolumn{1}{c}{UrQMD }\\ \hline
			  & \multicolumn{1}{c}{0.9}& $ (1.12\pm 0.09\pm 0.09)\times 10^{-4}  $ & $ (0.96 \pm 0.05)\times 10^{-4} $
			  \\
		$ d $	& \multicolumn{1}{c}{2.76}  & $ (1.53\pm 0.05\pm 0.13)\times 10^{-4}  $ &  $ (1.47 \pm 0.06)\times 10^{-4} $\\
			    & \multicolumn{1}{c}{7} & $ (2.02\pm 0.02\pm 0.17)\times 10^{-4}  $ &  $ (2.05 \pm 0.09) \times 10^{-4} $\\
			    & \multicolumn{1}{c}{0.9} & $ (1.11\pm 0.10\pm 0.09)\times 10^{-4}  $ &  $( 1.00 \pm 0.05)\times 10^{-4} $ \\
		$\overline{d}$ & \multicolumn{1}{c}{2.76} & $ (1.37\pm 0.04\pm 0.12)\times 10^{-4}  $ & $ (1.55 \pm 0.07)\times 10^{-4} $ \\
			   & \multicolumn{1}{c}{7} & $ (1.92\pm 0.02\pm 0.15)\times 10^{-4}  $ & $ (2.22 \pm 0.09) \times 10^{-4} $\\ 
               \hline
		   
		\end{tabular}
\caption{The integrated yield $(dN/dy)$ of deuterons and anti-deuterons in pp collisions with midrapidity $|y|<0.5$ at different center of mass energies as $\sqrt{s_{NN}}= 0.9, 2.76$ and $7$ TeV.} 
\end{table}
\begin{figure}[t]	
	\includegraphics[width=0.5\textwidth]{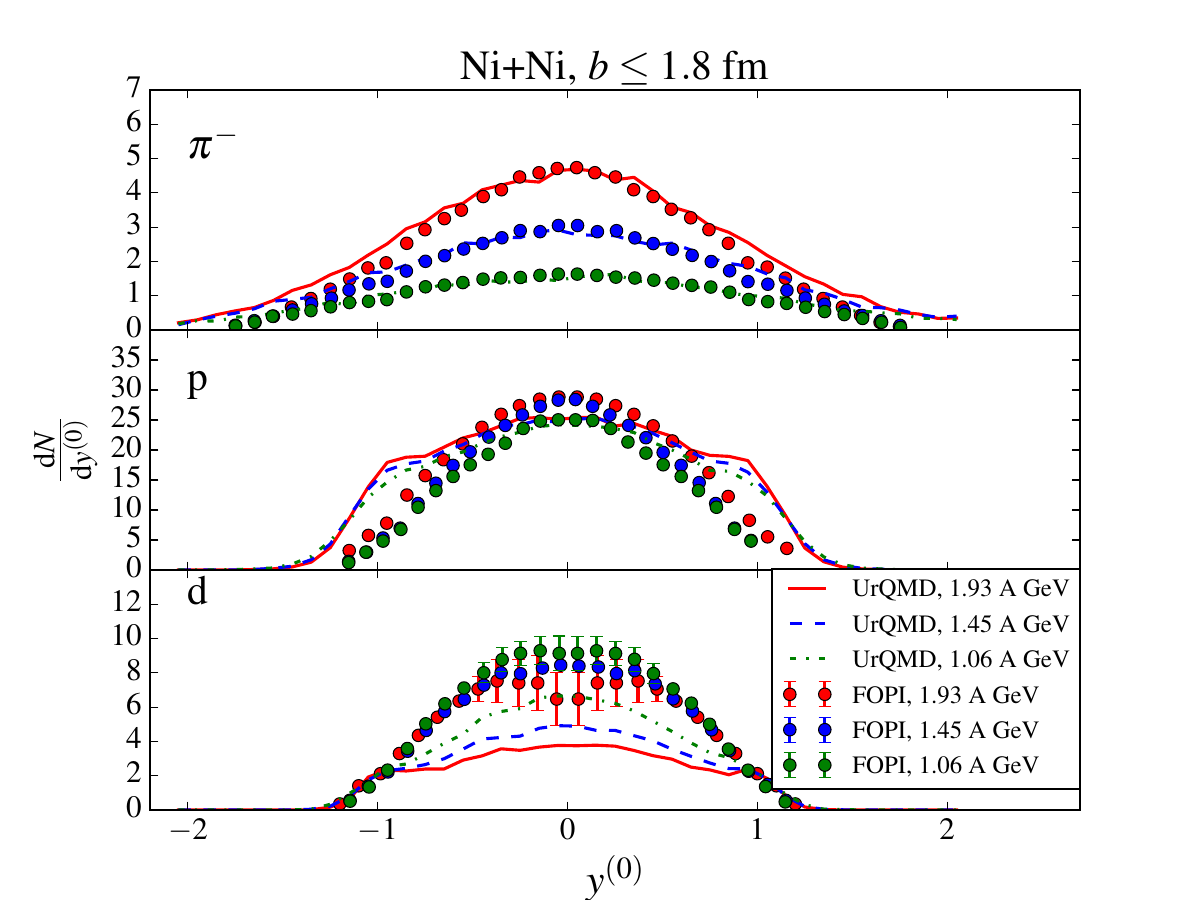}
	\caption{[Color online] The $ dN/dy^{(0)} $ distributions of deuterons, protons and $\pi^{-}$ for Ni+Ni collisions with $b\leq 1.8$ fm at beam energies $1.93$A, $1.45$A and $1.06$A GeV from our UrQMD simulation (lines) compared to the FOPI experimental data (symbols) \cite{Hong:1997mr}.
	}\label{f3}
\end{figure}		

Using the yields from Table I one can calculate the ratios of deuteron to proton (d/p) and anti-deuteron to anti-proton ($\overline{\mathrm{d}} $/ $ \overline{\mathrm{p}} $) as a function of energies $\sqrt{s_{NN}}= 53, 900, 2760$ and $7000$~GeV, as shown in Figure \ref{f2}. The open symbols are calculations by the UrQMD model and are compared to the experimental data. We find that at high energies, our results are consistent with the experimental data.

   \subsection{Nucleus-Nucleus reactions}
In the following we will present results of (anti-) deuteron production for collisions of light to heavy nuclei at various beam energies. 
\begin{figure}[t]	
	\includegraphics[width=0.5\textwidth]{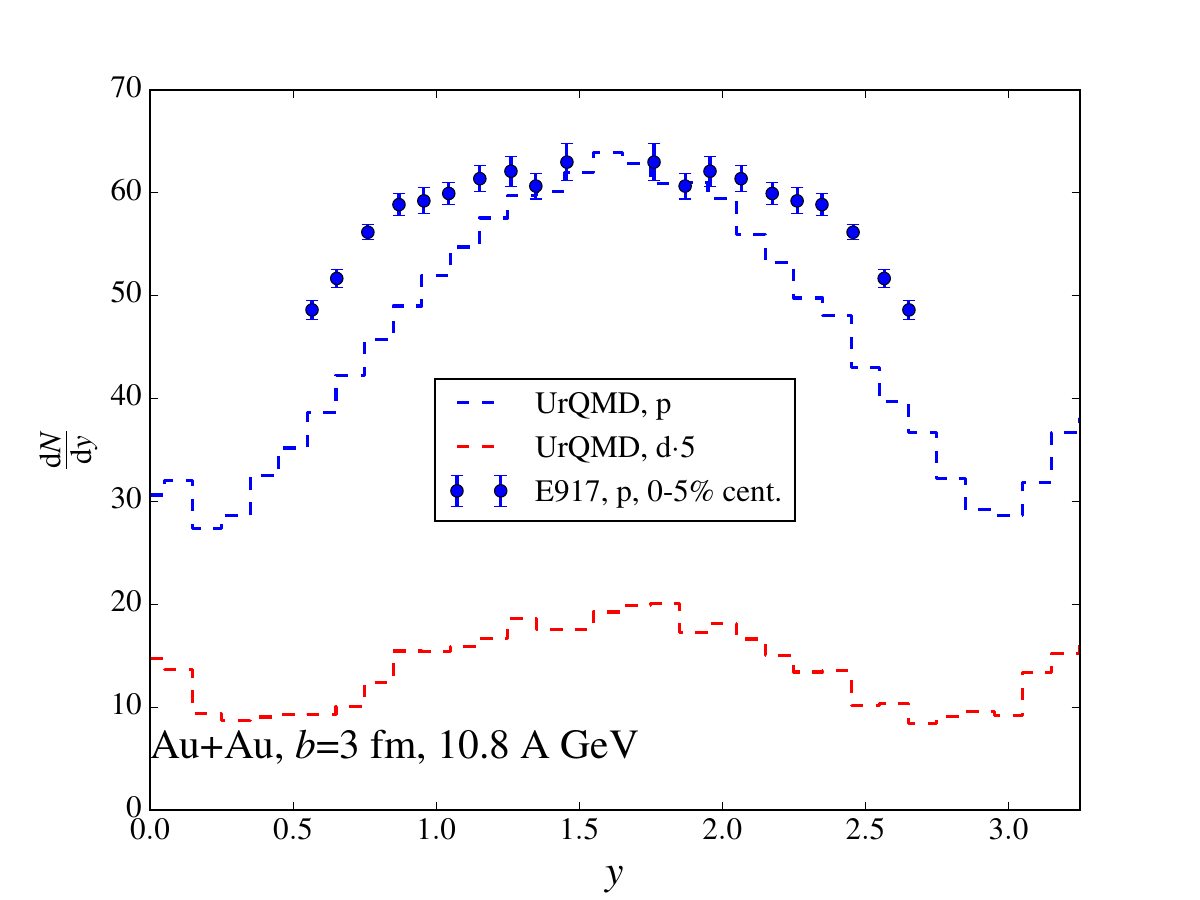}
	\caption{[Color online] Rapidity distributions of protons and deuterons (times 5) in Au+Au collisions at a beam energy of $10.8$A GeV with $b = 3$~fm. Results of the present UrQMD study (dashed lines) are compared to data from the E917 collaboration~\cite{Back:2000ru}.}\label{f4}
\end{figure}		
Starting with the lowest beam energies, we compare the rapidity distributions ($ dN/dy^{(0)} $) of deuterons, protons and $\pi^{-}$ for central Ni+Ni collisions (with $b\leq 1.8$ fm) at beam energies $1.93$A, $1.45$A and $1.06$A GeV with FOPI data \cite{Hong:1997mr}, as shown in Figure \ref{f3}. Here $y^{(0)}=y/y_{cm}$ is the rapidity scaled with the center-of-mass rapidity $ y_{cm} $. We find that our results are consistent with the FOPI data. For these low beam energies we observe deviations at forward and backward rapidities for the deuteron yields. However, these deviations are expected for the lowest beam energies as we did not include the effects of nuclear potential interactions in our simulations. These interactions are know to significantly effect the formation of nuclei at the lowest beam energies. As a result, the proton and deuteron rapidity distributions from UrQMD have a similar shape at different energies.

\begin{figure}[t]	
	\includegraphics[width=0.5\textwidth]{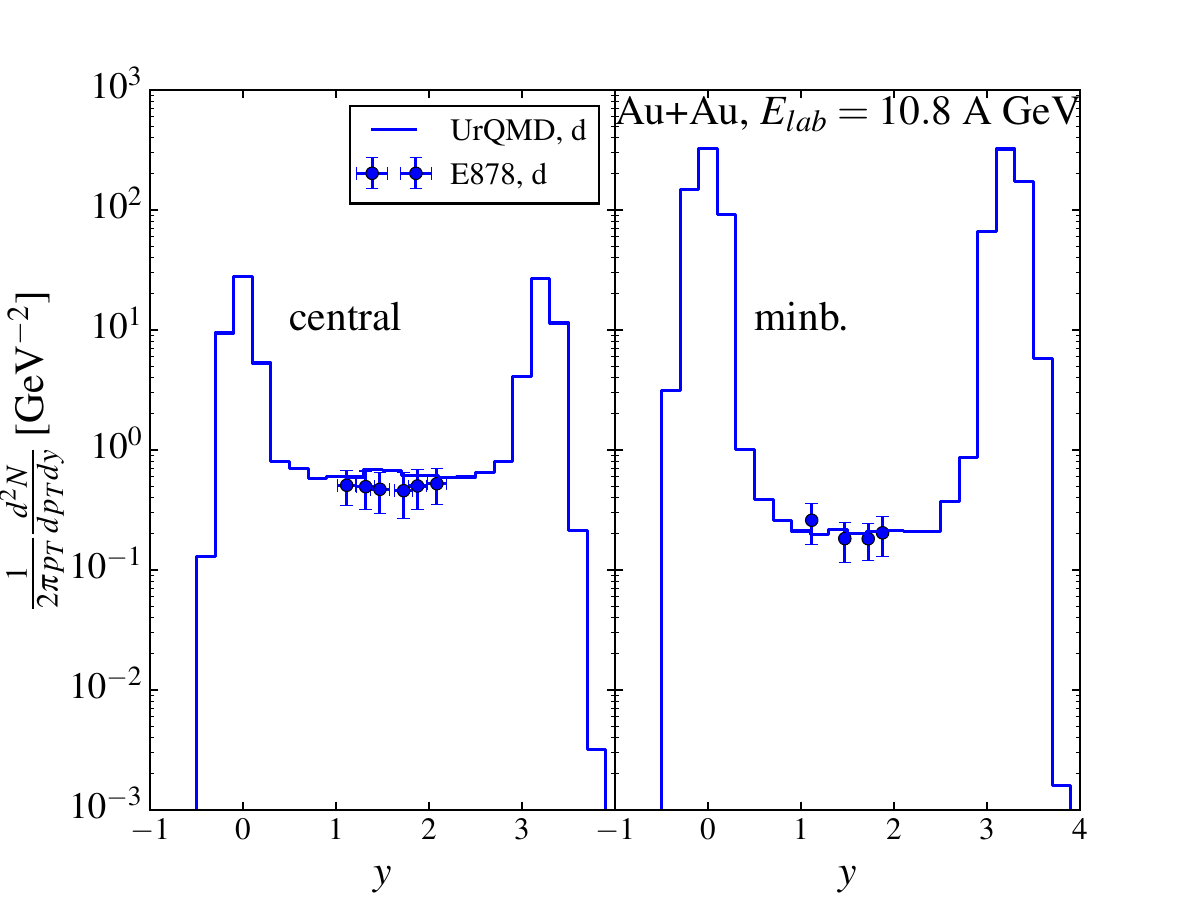}
	\caption{[Color online] Invariant yields of deuterons at $p_{t} = 0 $ as a function of rapidity in central (left) and minimum-bias (right) Au+Au collisions at a beam energy of $10.8$A GeV. Data from the E878 Experiment \cite{Bennett:1998be} are shown as symbols and the model calculations as lines.
	}\label{f5}
\end{figure}

Going to higher beam energies, Figures \ref{f4} and \ref{f5} present results of our calculations for Au+Au collisions at a beam energy of $10.8$A GeV.
The integrated rapidity distributions ($dN/dy$) of protons and deuterons in transport simulations for central collisions are shown in Figure \ref{f4}. Both, the proton and deuteron rapidity distributions are peaked around mid-rapidity (in the lab-frame), due to the increased stopping power in the symmetric heavy-ion collision.

Figure \ref{f5} shows a comparison of the calculations with experimental data at low transverse momentum. The invariant yields of deuterons at $p_{t} = 0 $ as a function of rapidity are shown for in central (left) and minimum-bias (right) Au+Au collisions at a beam energy of $10.8$A GeV. The lines show the UrQMD model calculations which are compared to data of the E878 experiment \cite{Bennett:1998be}. We find that our calculations are consistent with the E878 data. As expected, the yields in central collisions are higher than in minimum-bias collisions, due to the increased stopping in central collisions.

Finally, we present results for deuteron production in different colliding systems at a beam energy of $14.6$A GeV. At this beam energy a wealth of deuteron measurements were taken which allow a systematic comparison with our model calculations.
   
\begin{figure}[t]	
  	\includegraphics[width=0.5\textwidth]{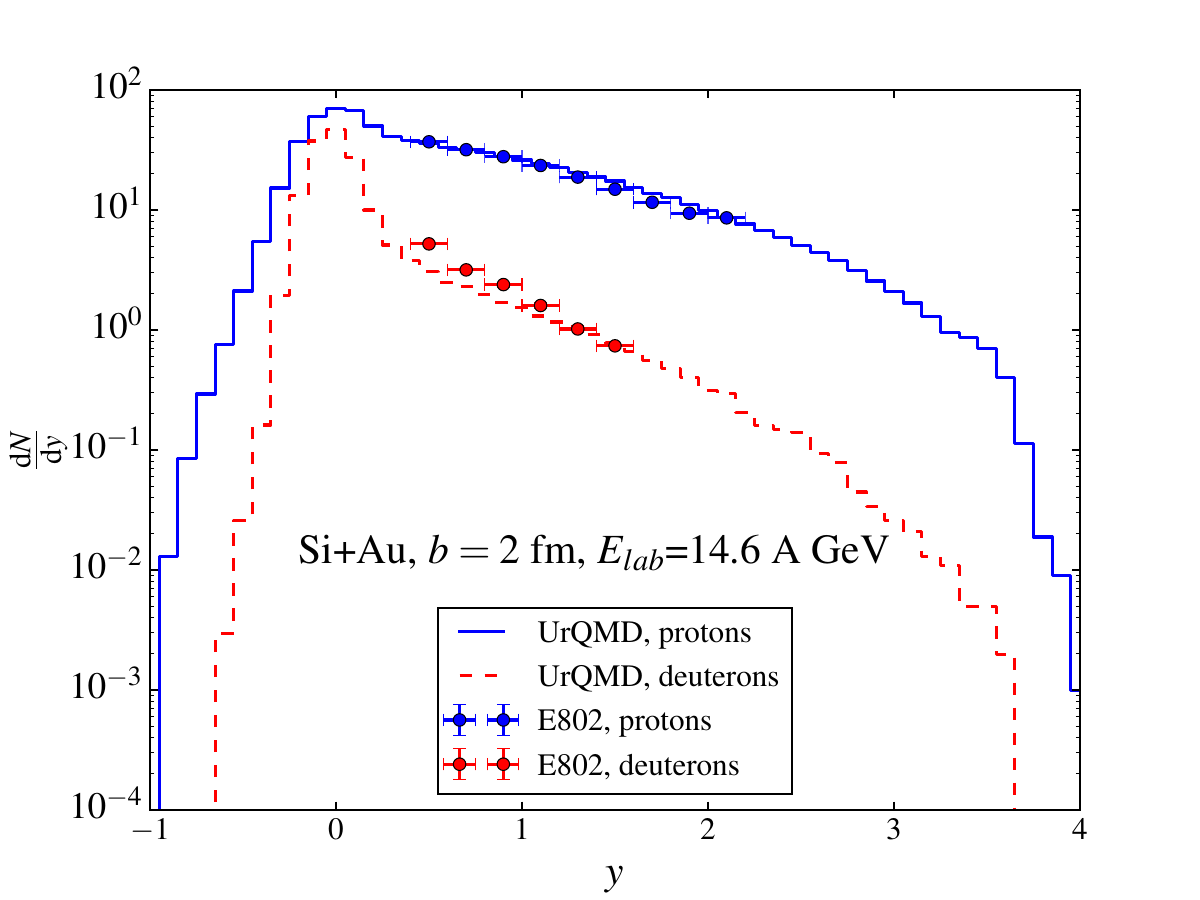}
  	\caption{[Color online] Rapidity distributions of protons and deuterons in Si+Au collisions at a beam energy of $14.6$A GeV with impact parameter $b = 2$~fm, comparing UrQMD results (lines) to data of E802 (symbols) \cite{Abbott:1994np}.
  	}\label{f6}
  \end{figure}

\begin{figure}[t]	
	\includegraphics[width=0.5\textwidth]{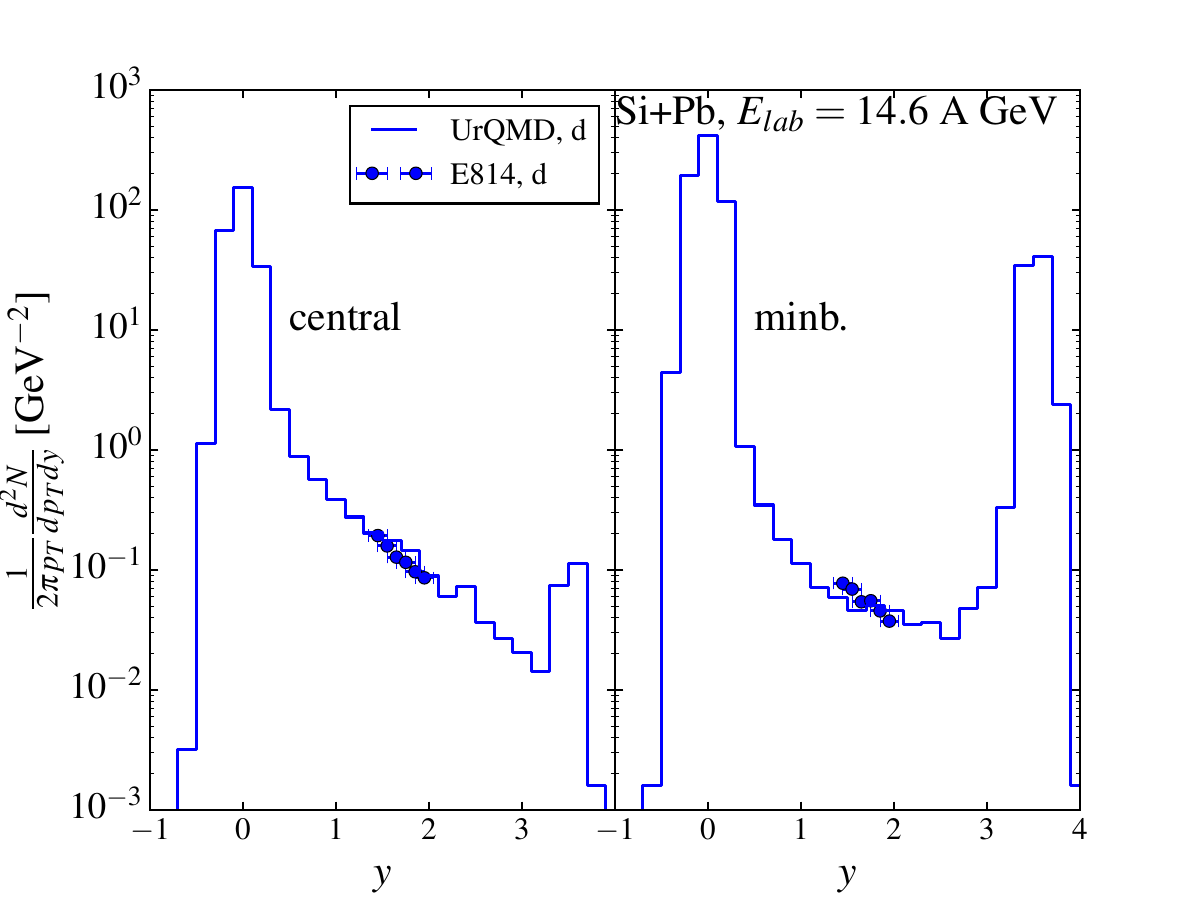}
	\caption{[Color online] Invariant yields of deuterons at $p_{t} = 0 $ as a function of rapidity in central (left) and minimum-bias (right) Si+Pb collisions at a beam energy of $14.6$A GeV. Data of the  E814 Experiment \cite{Barrette:1994tw} are shown as symbols and the model calculations as lines.
	}\label{f7}
\end{figure}

First, shown in Figures \ref{f6} and \ref{f7} are the rapidity distributions of protons and deuterons in Si+Au and Si+Pb collisions at a beam energy of $14.6$A GeV for different centralities. Again we observe that the rapidity distributions of protons and deuterons are in good agreement with the experimental data of the E802\cite{Abbott:1994np} collaboration as shown in Figure \ref{f6}. 
  
Figure \ref{f7} shows invariant yields of deuterons as a function of rapidity in central (left) and minimum-bias (right) Si+Pb collisions at a beam energy of $14.6$A GeV at $p_{t} = 0 $. The lines indicate the UrQMD calculations, the symbols denote the E814 data from Ref.~\cite{Barrette:1994tw}. We find that the calculated invariant yields are in good agreement with the measured E814 data. 

\begin{figure}[t]	
	\includegraphics[width=0.5\textwidth]{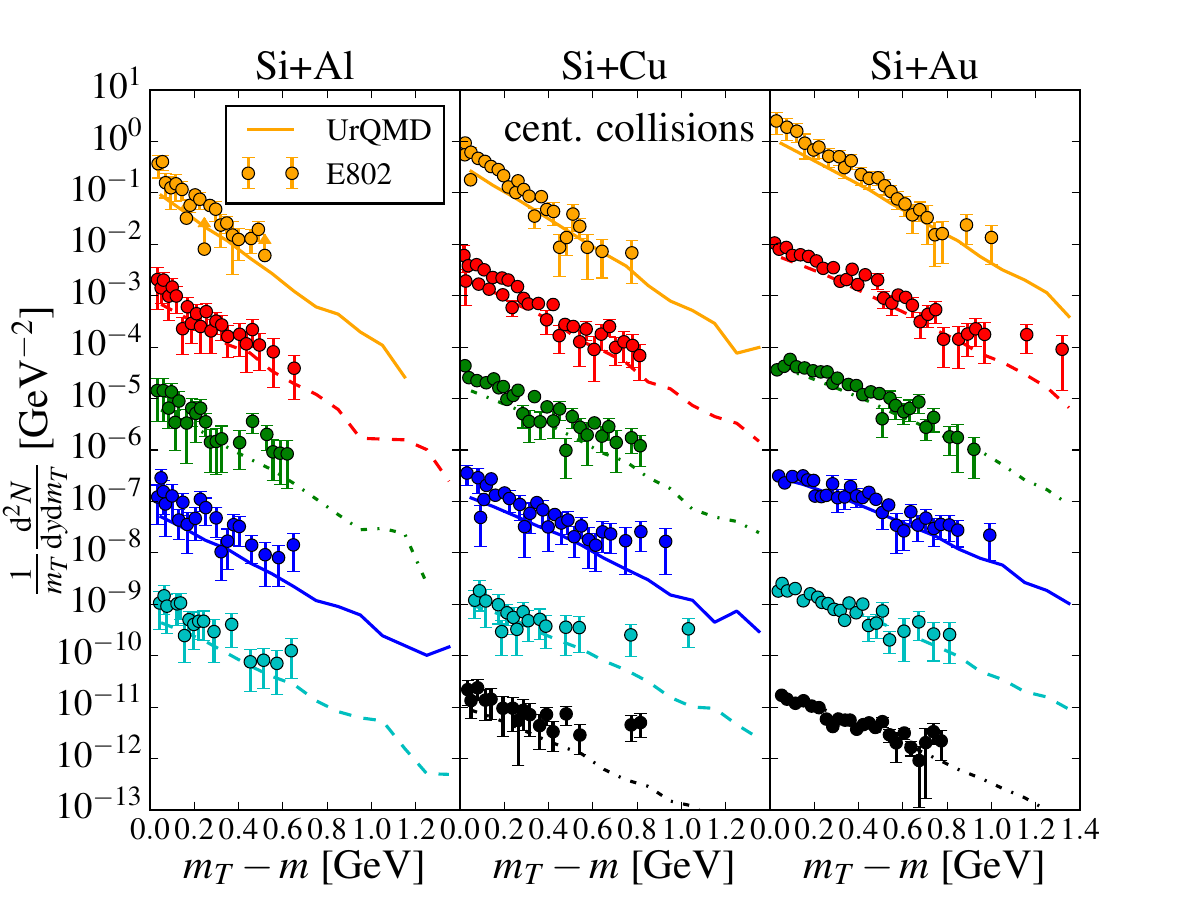}
    	\caption{[Color online] Invariant yields of deuterons as a function of $ m_{t}-m $ in central Si+Al, Si+Cu and Si+Au collisions at a beam energy of $14.6$A GeV. The rapidity interval is $y = 0.5$ to $1.5$ with $ \Delta y = 0.2 $.  Each successive spectrum is divided by 100 for visual clarity. The symbols denote data of the E802 collaboration \cite{Abbott:1994np}.
	}\label{f8}
\end{figure}

Moreover, we show invariant yields of deuterons as a function of $ m_{t}-m $ in central Si+Al, Si+Cu and Si+Au collisions 
at a beam energy of $14.6$A GeV. In Figure \ref{f8} we compare our results to data of the experiment E802 \cite{Abbott:1994np}. For central collisions, the rapidity intervals re $y = 0.5$ to $1.5$ with $ \Delta y = 0.2 $. Each successive spectrum is divided by 100 for better visibility. The invariant yields are determined as 
\begin{equation}
E\left( \frac{d^{3}N}{dp^{3}}\right)= \left(\frac{1}{2 \pi m_{t}}\right) \left( \frac{d^{2}N}{dydm_{t}}\right)
\end{equation}
where $ m_{t} $ is the transverse mass 
\begin{equation}
m_{t}=(p_{t}^{2}+m^{2})^{1/2}~,
\end{equation}
and $E$ is the energy and $p$ the momentum.

We find that our results are consistent with the data from the experiment E802. 
The resulting invariant yields of deuterons for the three targets and for each rapidity interval show that the invariant yields decrease with increasing rapidity until the fragmentation region.

Going up in energy, we next explore the CERN-SPS energy regime. The NA49 experiment explored deuteron formation in great detail at various energies and centralities. The data of the NA49 experiment will be compared to UrQMD calculations for Pb+Pb collisions at different energies.
\begin{figure}[t]	
	\includegraphics[width=0.5\textwidth]{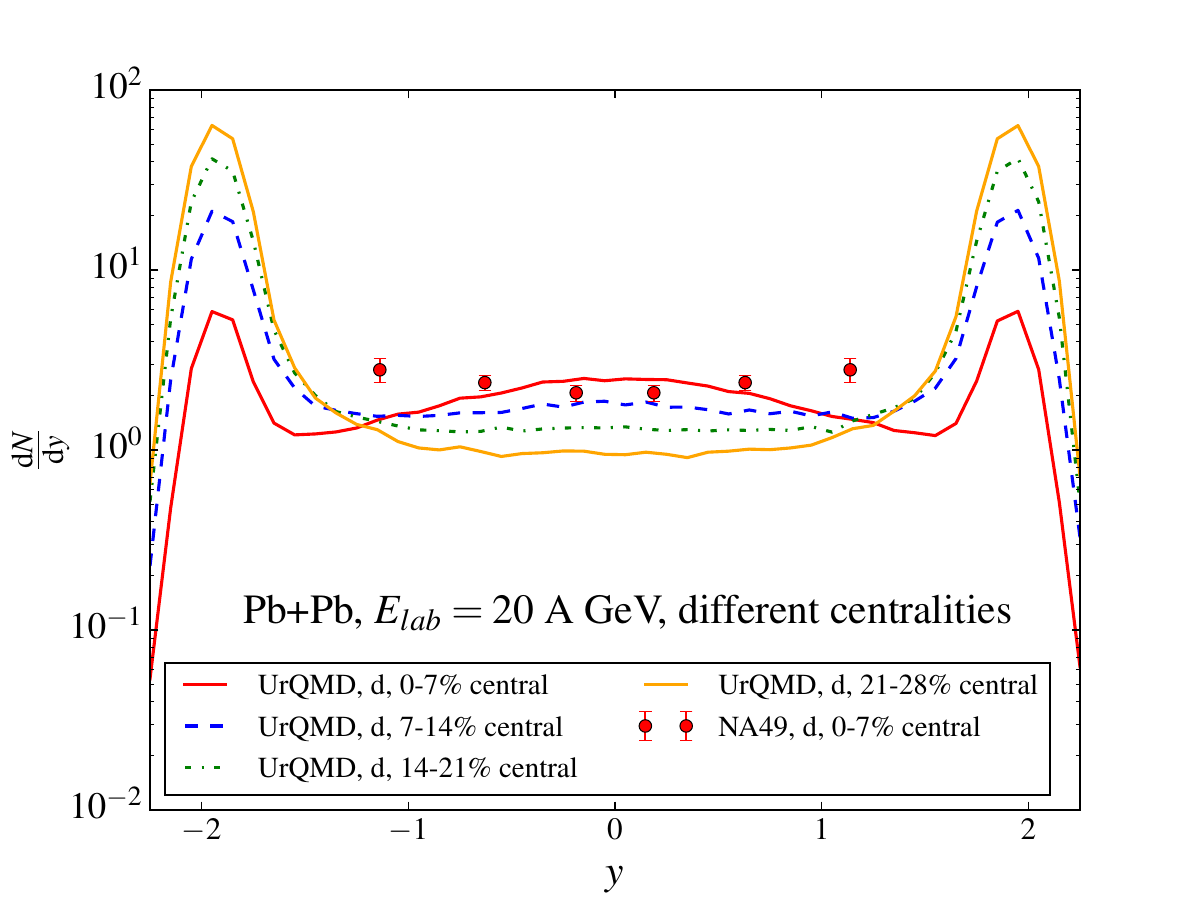}
	\caption{[Color online] Deuteron yields as a function of rapidity in Pb+Pb collisions at a beam energy of $20$A GeV for different centralities. The symbols denote data of the NA49 experiment~\cite{Anticic:2016ckv}.}\label{fsps1}
\end{figure}		
\begin{figure}[t]	
	\includegraphics[width=0.5\textwidth]{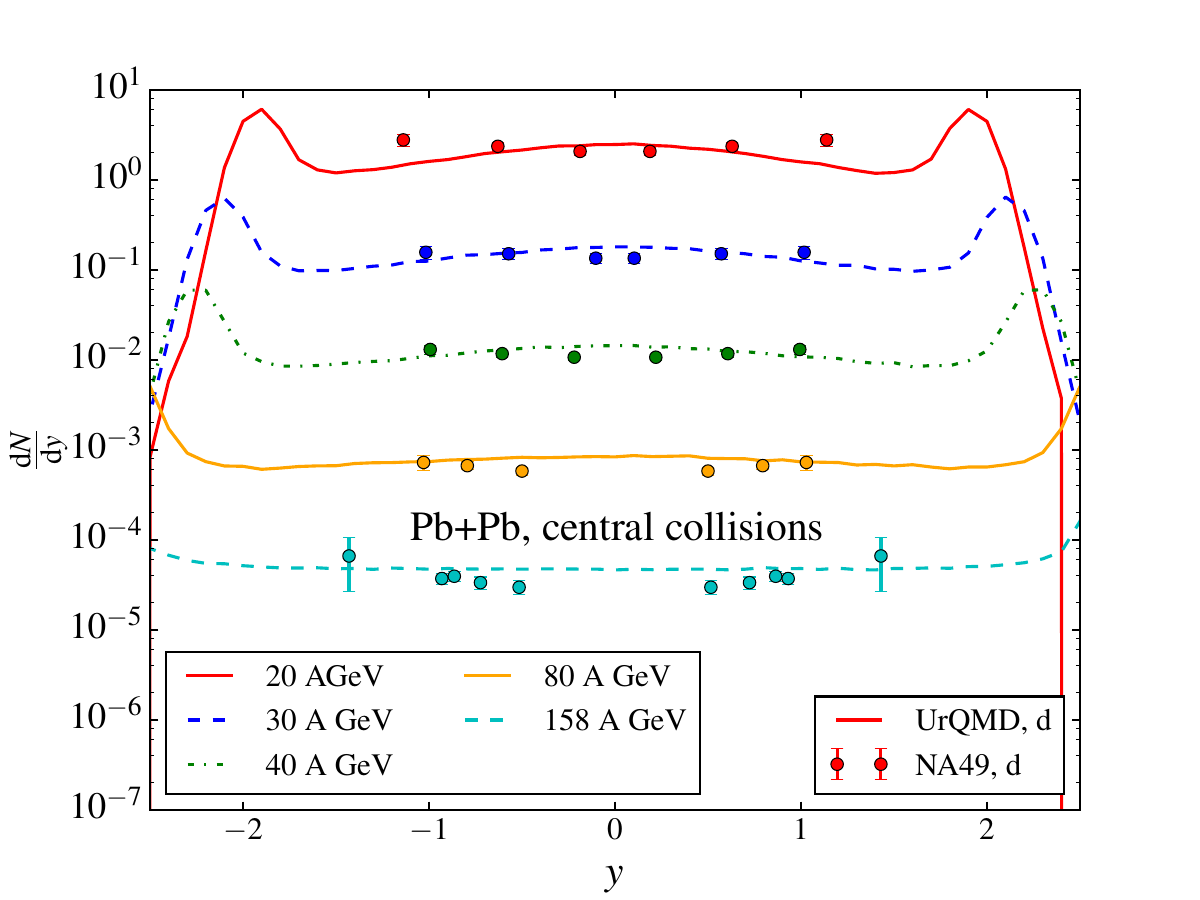}
	\caption{[Color online] Rapidity dependence of deuteron yields for various beam energies in Pb+Pb collisions for different centralities. The symbols denote data of the NA49 experiment~\cite{Anticic:2016ckv}, the lines show the calculations. Each spectrum is successively divided by a factor of 10.}\label{fsps2}
\end{figure}		%
\begin{figure}[t]	
	\includegraphics[width=0.5\textwidth]{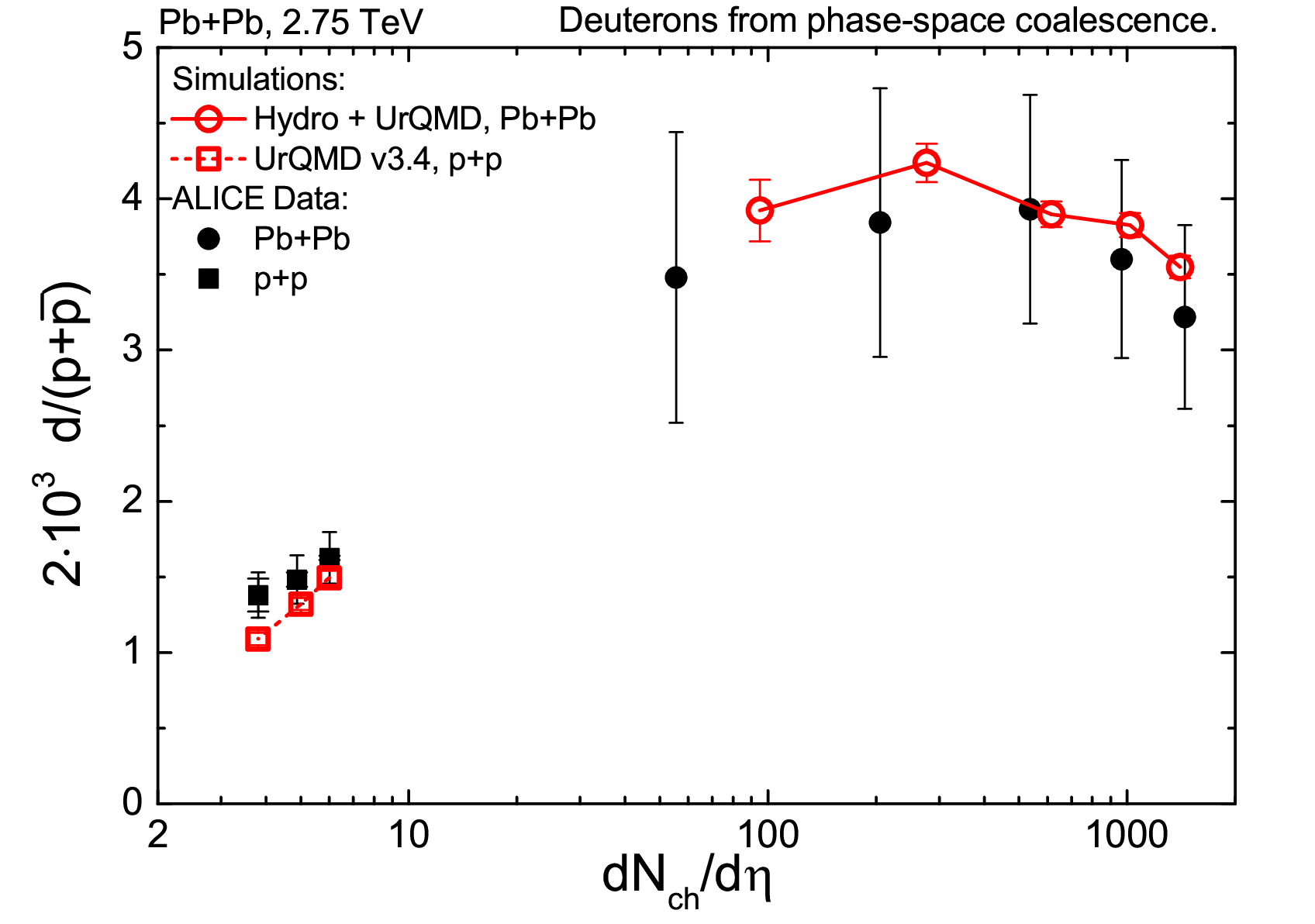}
	\caption{[Color online] Ratio of deuteron to protons+anti-protons in Pb+Pb collisions at $\sqrt{s_{NN}}=2.76$ TeV as a function of the charged particle multiplicity. In addition the values for proton-proton reaction at various energies are also indicated on the left part of the figure. The UrQMD results are compare to ALICE data \cite{Adam:2015vda,Acharya:2017fvb}. }\label{f8b}
\end{figure}		
Figure \ref{fsps1} shows the deuteron multiplicity as a function of rapidity for Pb+Pb collisions at a beam energy of 20A GeV for different centralities. The lines denote the UrQMD calculations and the symbols denote the data of the NA49 collaboration~\cite{Anticic:2016ckv}. The calculations are in good agreement with the experimental data. However, one can observe a small deviation to the experimental data which is due to a stronger baryon stopping in UrQMD as compared to the NA49 data, when going towards more central collisions.

Figure \ref{fsps2} shows the deuteron multiplicity as a function of rapidity at  beam energies of 20A GeV, 30A GeV, 40A GeV, 80A GeV and 158A GeV for central Pb+Pb collisions. The lines denote the UrQMD calculations and the symbols the experimental data of the NA49 collaboration~\cite{Anticic:2016ckv}.  For visibility the calculations and the data are divided by a factor of $10$ successively. Also here, the calculations are in good agreement with the experimental data.
Given the results presented above, we have established that deuteron production at moderate energies can be very well described by a single energy and system size independent phase space coalescence parameter set. 

In the last steps we want to explore, if this single parameter set can also be used to describe deuteron production at the highest available energies, namely Pb+Pb collisions at the LHC. In Fig. \ref{f8b} we show the 
ratio of deuterons to protons plus anti-protons in Pb+Pb collisions at $\sqrt{s_{NN}}=2.76$ TeV as a function of the charged particle multiplicity. Here we used the UrQMD+hydro hybrid version of the model to properly take into account the long hydrodynamical expansion of the fireball. The coalescence procedure is applied after the hadronic rescattering phase, as described above. In addition we also show the values for proton-proton reactions at different beam energies with their corresponding $N_{\mathrm{ch}}$, indicated as open squares in the left part of the figure. The UrQMD/hybrid results are compared to ALICE data \cite{Anielski:2015zna}. One observes a very good agreement between the measured data and the calculations over the whole range of centralities/multiplicities. Thus, we can conclude that deuteron production at the LHC can be very well described by coalescence of protons and neutrons with the same parameters used at lower collision energies.

\begin{figure}[t]	
	\includegraphics[width=0.5\textwidth]{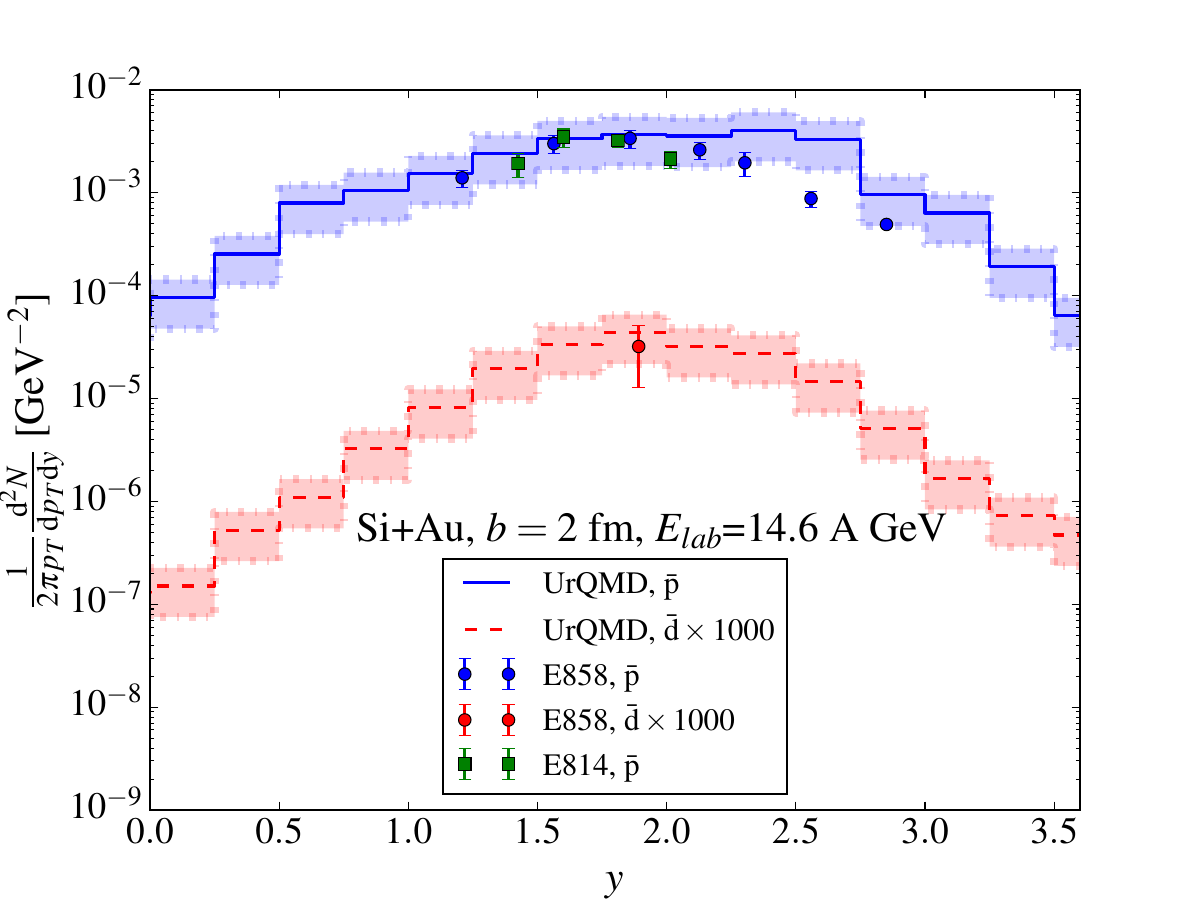}
	\caption{[Color online] Invariant yields of anti-deuterons ($ \overline{d} $) and anti-protons ($ \overline{p} $) at $p_{t} = 0 $ as a function of rapidity in minimum-bias Si+Au collisions at a beam energy of $14.6$A GeV. The symbols denote data of the experiments E814 \cite{Barrette:1993hc} and E858 \cite{Kumar:1994rp} and the histograms show results from the UrQMD model.	}\label{f9}
\end{figure}

Finally we present the invariant yields of anti-deuterons ($ \overline{d} $) and anti-protons ($ \overline{p} $) at $p_{t} = 0 $ as a function of rapidity in minimum-bias Si+Au collisions at a beam energy of $14.6$A GeV. Our results are compared to the data from the E814~\cite{Barrette:1993hc} and E858~\cite{Kumar:1994rp} experiments shown in Figure \ref{f9}. One observes that the UrQMD model results are in good agreement with the experimental data.

\subsection{ Excitation function}

\begin{figure}[t]	
	\includegraphics[width=0.5\textwidth]{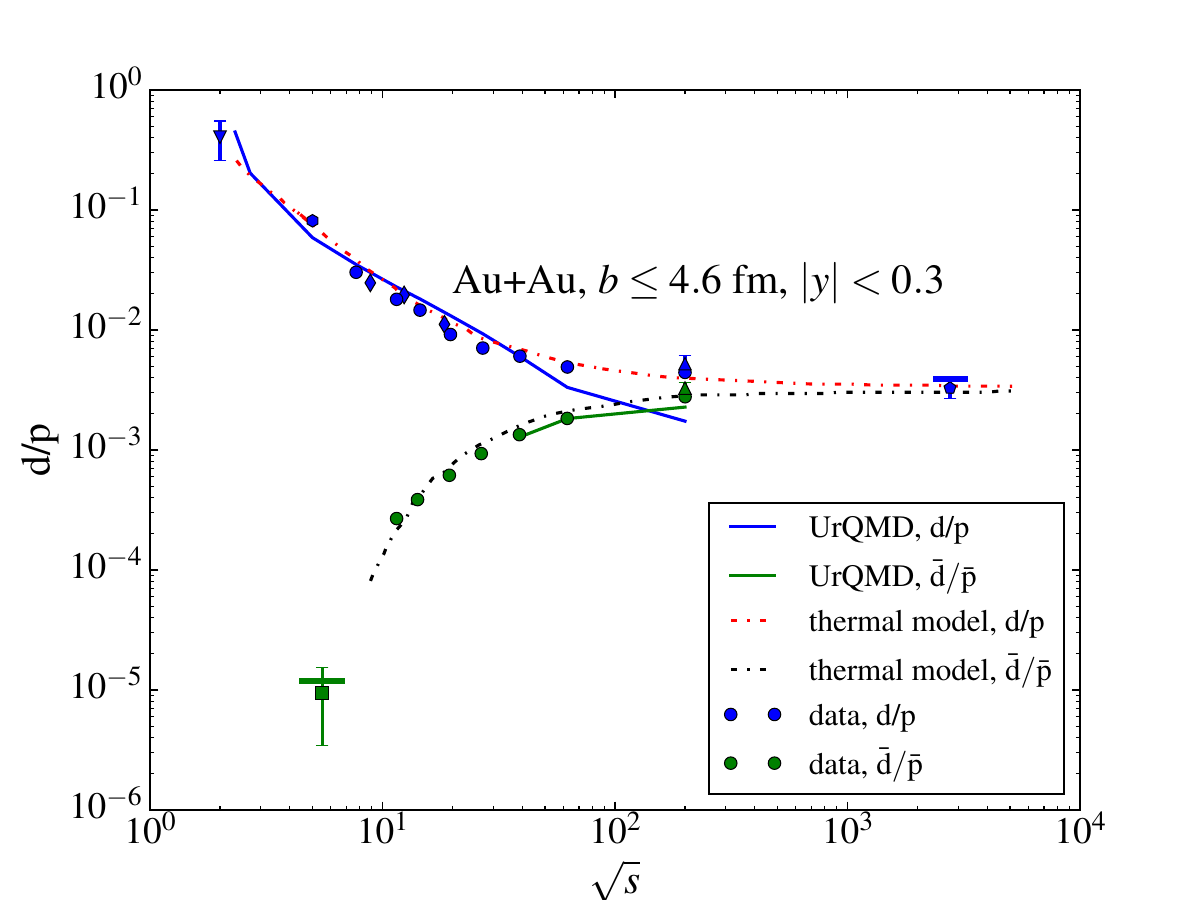}
	\caption{[Color online] Energy dependence of the deuteron to proton ratio in Au+Au collisions with $b \leq 4.6$~fm and $|y|<0.3$ at $ \sqrt{s_{NN}}=2, 5, 7.7, 11.5, 14.5, 17$ and $19.6$~GeV. The solid lines represent UrQMD model results, the dotted lines denote the thermal model fit \cite{Andronic:2010qu} and the symbols denote experimental data from various collaborations (triangle down: SIS~\cite{Cleymans:1998yb}, hexagon: E802~\cite{Ahle:1999in}, triangles up: PHENIX~\cite{Adler:2004uy}, blue diamonds: NA49~\cite{Anticic:2016ckv}, circles: STAR~\cite{Yu:2017bxv}, pentagon: ALICE~\cite{Anielski:2015zna}, square: E814~\cite{Barrette:1993hc}). The blue horizontal line represents the UrQMD+hydro result on the d/p ratio at 2.76 TeV and the green horizontal line represents the UrQMD result of the $\bar{\mathrm{d}}/\bar{\mathrm{p}}$ ratio in Si+Au collisions at $E_{lab}=14.6$A GeV.}\label{f10}
\end{figure}

In the last section we summarize the energy dependence of the deuteron-to-proton ratio (and anti-deuteron to anti-proton ratio) for central Au+Au collisions. The mid-rapidity ratios ($|y|<0.3$) are calculated at $\sqrt{s_{NN}}=2, 5, 7.7, 11.5,  14.5, 17$ and $19.6$~GeV and  are shown as solid lines in Figure \ref{f10}. The model calculations are  compared to a thermal model fit (dotted line) \cite{Andronic:2010qu} and experimental data from the SIS~\cite{Cleymans:1998yb}, E802~\cite{Ahle:1999in}, PHENIX~\cite{Adler:2004uy}, NA49~\cite{Anticic:2016ckv}, STAR~\cite{Yu:2017bxv}, ALICE~\cite{Anielski:2015zna}, and E814 collaboration~\cite{Barrette:1993hc}. We find that our results are consistent with both the thermal model and the experimental data. However, it should be noted that the cascade mode calculations of UrQMD starts to underestimate the $d/p$ ratio above 80 GeV center-of-mass energy. Above this energy, hybrid model calculations are necessary as indicated by the LHC result.  The decrease of the d/p ratio towards higher energies is due to the decreasing phase space density for baryons at higher beam energies. The decrease of the $\bar{\mathrm{d}}/\bar{\mathrm{p}}$ ratio towards lower energies indicates the surface freeze-out of the anti-protons and thus also a decreased phase space density \cite{Mrowczynski:1993cx,Bleicher:1995dw}.

\section{Summary}
In the present paper, we have provided a benchmarking transport model study of deuteron production in the FAIR and up to the LHC energy regime. The UrQMD+coalescence model with parameters $\Delta p_{max}=0.285$~GeV/c and $\Delta r_{max}= 3.575$~fm provides a very good description of the available data of deuteron production. Starting from proton induced reactions (p+p and p+Au, p+Be) at low and high energies towards Pb+Pb reaction at the CERN-SPS and CERN-LHC we find that deuteron production for all systems can be described by coalescence with the same phase space parameters. Given the current discussion of the deuteron production at LHC, we want to stress that coalescence provides similar results for the d/p ratios as the thermal model over the whole range of expected energies. In addition it captures the decrease of the d/p ratio for the high centrality bin in Pb+Pb reactions at the LHC. Moreover, we have demonstrated that the data of invariant yields of ($ \overline{d} $) and ($ \overline{p} $) at $p_{t} = 0 $ as a function of rapidity in minimum-bias Si+Au collisions at $14.6$A GeV is in agreement with the UrQMD calculations. 

\section{Acknowledgments}
This work was supported by Thailand Research Fund (TRF-RGJ PHD/0185/2558), Deutscher Akademischer Austauschdienst (DAAD), HIC for FAIR and in the framework of COST Action CA15213 THOR. The computational resources have been provided by the LOEWE Frankfurt Center for Scientific Computing (LOEWE-CSC), the FUCHS-CSC, and the National e-Science Infrastructure Consortium, Thailand. KT acknowledges the financial support from Naresuan University R2560C185. AL, CH and YY acknowledge support from SUT-CHE-NRU project of Thailand.


\begin{thebibliography}{100}
 \bibitem{Andronic:2012dm} 
  A.~Andronic, P.~Braun-Munzinger, K.~Redlich and J.~Stachel,
  Nucl.\ Phys.\ A {\bf 904-905}, 535c (2013)
  doi:10.1016/j.nuclphysa.2013.02.070
  \bibitem{Mrowczynski:2016xqm} 
  S.~Mrowczynski,
  Acta Phys.\ Polon.\ B {\bf 48}, 707 (2017)
  doi:10.5506/APhysPolB.48.707

\bibitem{Mrowczynski:1992gc} 
  S.~Mrowczynski,
  Phys.\ Lett.\ B {\bf 277}, 43 (1992).
  doi:10.1016/0370-2693(92)90954-3
  \bibitem{Bass:1998ca} 
  S.~A.~Bass {\it et al.},
  Prog.\ Part.\ Nucl.\ Phys.\  {\bf 41}, 255 (1998).
  \bibitem{Bleicher:1999xi} 
  M.~Bleicher {\it et al.},
  J.\ Phys.\ G {\bf 25}, 1859 (1999).
\bibitem{Graef:2014mra} 
  G.~Graef, J.~Steinheimer, F.~Li and M.~Bleicher,
  Phys.\ Rev.\ C {\bf 90}, 064909 (2014)
\bibitem{Olive:2016xmw} 
C.~Patrignani {\it et al.} [Particle Data Group],
Chin.\ Phys.\ C {\bf 40}, no. 10, 100001 (2016).
\bibitem{Sato:1981ez} 
  H.~Sato and K.~Yazaki,
  Phys.\ Lett.\  {\bf 98B}, 153 (1981).
  doi:10.1016/0370-2693(81)90976-X
 \bibitem{Steinheimer:2012tb} 
 J.~Steinheimer, K.~Gudima, A.~Botvina, I.~Mishustin, M.~Bleicher and H.~St\"ocker,
 Phys.\ Lett.\ B {\bf 714}, 85 (2012)
 doi:10.1016/j.physletb.2012.06.069
 \bibitem{Gyulassy:1982pe} 
 M.~Gyulassy, K.~Frankel and E.~a.~Remler,
 Nucl.\ Phys.\ A {\bf 402}, 596 (1983).
 doi:10.1016/0375-9474(83)90222-1
 \bibitem{Aichelin:1987rh} 
  J.~Aichelin and E.~A.~Remler,
  Phys.\ Rev.\ C {\bf 35}, 1291 (1987).
  doi:10.1103/PhysRevC.35.1291
 \bibitem{Nagle:1996vp} 
 J.~L.~Nagle, B.~S.~Kumar, D.~Kusnezov, H.~Sorge and R.~Mattiello,
 Phys.\ Rev.\ C {\bf 53}, 367 (1996).
 doi:10.1103/PhysRevC.53.367
 \bibitem{Nagle:1994wj} 
 J.~L.~Nagle, B.~S.~Kumar, M.~J.~Bennett, S.~D.~Coe, G.~E.~Diebold, J.~K.~Pope, A.~Jahns and H.~Sorge,
 Phys.\ Rev.\ Lett.\  {\bf 73}, 2417 (1994).
 doi:10.1103/PhysRevLett.73.2417
\bibitem{Xia:2014rua} 
  Y.~Xia, J.~Xu, B.~A.~Li and W.~Q.~Shen,
  Nucl.\ Phys.\ A {\bf 955}, 41 (2016)
  doi:10.1016/j.nuclphysa.2016.06.001
  [arXiv:1411.3057 [nucl-th]].
 \bibitem{Ko:2010zza} 
  C.~M.~Ko, Z.~W.~Lin and Y.~Oh,
  Nucl.\ Phys.\ A {\bf 834}, 253C (2010).
  doi:10.1016/j.nuclphysa.2009.12.052
 \bibitem{Zhu:2015voa} 
 L.~Zhu, C.~M.~Ko and X.~Yin,
 Phys.\ Rev.\ C {\bf 92}, no. 6, 064911 (2015)
 doi:10.1103/PhysRevC.92.064911
 \bibitem{Li:2015pta}
  Q.~Li, Y.~Wang, X.~Wang, C.~Shen and M.~Bleicher,
  arXiv:1507.06033 [hep-ph].
\bibitem{Abbott:1991en} 
  T.~Abbott {\it et al.} [E-802 Collaboration],
  Phys.\ Rev.\ D {\bf 45}, 3906 (1992).
\bibitem{Alper:1973my} 
  B.~Alper {\it et al.},
  Phys.\ Lett.\  {\bf 46B}, 265 (1973).
  doi:10.1016/0370-2693(73)90700-4
\bibitem{Henning:1977mt} 
  S.~Henning {\it et al.} [British-Scandinavian-MIT Collaboration],
  Lett.\ Nuovo Cim.\  {\bf 21}, 189 (1978).
  doi:10.1007/BF02822248
\bibitem{Alper:1975jm} 
  B.~Alper {\it et al.} [British-Scandinavian Collaboration],
  Nucl.\ Phys.\ B {\bf 100}, 237 (1975).
  doi:10.1016/0550-3213(75)90618-5
\bibitem{Acharya:2017fvb} 
  S.~Acharya {\it et al.} [ALICE Collaboration],
  Phys.\ Rev.\ C {\bf 97}, no. 2, 024615 (2018)
  doi:10.1103/PhysRevC.97.024615

\bibitem{Hong:1997mr} 
B.~Hong {\it et al.} [FOPI Collaboration],
Phys.\ Rev.\ C {\bf 57}, 244 (1998)
Erratum: [Phys.\ Rev.\ C {\bf 58}, 603 (1998)]
\bibitem{Back:2000ru} 
  B.~B.~Back {\it et al.} [E917 Collaboration],
  Phys.\ Rev.\ Lett.\  {\bf 86}, 1970 (2001)
  doi:10.1103/PhysRevLett.86.1970
  [nucl-ex/0003007].

\bibitem{Bennett:1998be} 
M.~J.~Bennett {\it et al.} [E878 Collaboration],
Phys.\ Rev.\ C {\bf 58}, 1155 (1998).
doi:10.1103/PhysRevC.58.1155

\bibitem{Abbott:1994np} 
  T.~Abbott {\it et al.} [E-802 Collaboration],
  Phys.\ Rev.\ C {\bf 50}, 1024 (1994).
  doi:10.1103/PhysRevC.50.1024
\bibitem{Anticic:2016ckv} 
  T.~Anticic {\it et al.} [NA49 Collaboration],
  Phys.\ Rev.\ C {\bf 94}, no. 4, 044906 (2016)
  doi:10.1103/PhysRevC.94.044906
  [arXiv:1606.04234 [nucl-ex]].
\bibitem{Barrette:1994tw} 
J.~Barrette {\it et al.} [E814 Collaboration],
Phys.\ Rev.\ C {\bf 50}, 1077 (1994).
doi:10.1103/PhysRevC.50.1077
\bibitem{Anielski:2015zna} 
  J.~Anielski [ALICE Collaboration],
  J.\ Phys.\ Conf.\ Ser.\  {\bf 612}, no. 1, 012014 (2015).
  doi:10.1088/1742-6596/612/1/012014
 
\bibitem{Cleymans:1998yb} 
  J.~Cleymans, H.~Oeschler and K.~Redlich,
  Phys.\ Rev.\ C {\bf 59}, 1663 (1999)
  doi:10.1103/PhysRevC.59.1663
  [nucl-th/9809027].

\bibitem{Andronic:2010qu} 
  A.~Andronic, P.~Braun-Munzinger, J.~Stachel and H.~St\"ocker,
  Phys.\ Lett.\ B {\bf 697}, 203 (2011)
\bibitem{Yu:2017bxv} 
  N.~Yu [STAR Collaboration],
  Nucl.\ Phys.\ A {\bf 967}, 788 (2017)
  doi:10.1016/j.nuclphysa.2017.06.046
\bibitem{Ahle:1999in} 
  L.~Ahle {\it et al.} [E802 Collaboration],
  Phys.\ Rev.\ C {\bf 60}, 064901 (1999).
  doi:10.1103/PhysRevC.60.064901
\bibitem{Adler:2004uy} 
  S.~S.~Adler {\it et al.} [PHENIX Collaboration],
  Phys.\ Rev.\ Lett.\  {\bf 94}, 122302 (2005)
  doi:10.1103/PhysRevLett.94.122302
  [nucl-ex/0406004].
\bibitem{Barrette:1993hc} 
J.~Barrette {\it et al.} [E814 Collaboration],
Phys.\ Rev.\ Lett.\  {\bf 70}, 1763 (1993).
doi:10.1103/PhysRevLett.70.1763
\bibitem{Mrowczynski:1993cx} 
  S.~Mrowczynski,
  Phys.\ Lett.\ B {\bf 308}, 216 (1993).
  doi:10.1016/0370-2693(93)91274-Q
  \bibitem{Bleicher:1995dw} 
  M.~Bleicher, C.~Spieles, A.~Jahns, R.~Mattiello, H.~Sorge, H.~St\"ocker and W.~Greiner,
  Phys.\ Lett.\ B {\bf 361}, 10 (1995)
  doi:10.1016/0370-2693(95)01159-N
  [nucl-th/9506009].
\bibitem{Kumar:1994rp} 
B.~S.~Kumar {\it et al.} [E858/E878 Collaboration],
Nucl.\ Phys.\ A {\bf 566}, 439C (1994).
doi:10.1016/0375-9474(94)90664-5

\bibitem{Adam:2015vda} 
  J.~Adam {\it et al.} [ALICE Collaboration],
  Phys.\ Rev.\ C {\bf 93}, no. 2, 024917 (2016).
 
\end{thebibliography}
\end{document}